\documentclass[aps,12pt]{revtex4-2}

\usepackage{amsmath}
\usepackage{amssymb}
\usepackage{commath}
\usepackage{graphicx}
\usepackage{float}
\usepackage{ulem} 

\usepackage[toc,page]{appendix}

\usepackage{graphicx}
\usepackage{dcolumn}
\usepackage{bm}

\usepackage{color}

\begin{document}


\title{The Persistence of Nonlinear Gravitational Wave Memory}  
\author{Robert R. Caldwell} 

\affiliation{%
 Department of Physics and Astronomy, Dartmouth College, Hanover, NH 03755, USA}%
\date{\today}

\maketitle

{\bf 
Nonlinear gravitational wave memory is a surprise of theoretical physics. Whereas it is understood that a gravitational wave induces oscillatory squeezing and stretching motion in a collection of freely-falling test masses, it is unexpected that the wave leaves a residual displacement of the test masses. This displacement is the tribute \textit{\textbf in memoriam} to the passing wave. The memory originates in a nonlinear feature of gravitation. Whilst merging black holes are a significant source of gravitational waves, the gravitational wave energy itself is a further source of gravitational waves. The memory is often described as a permanent displacement of the test masses caused by a burst of primary gravitational waves. But as we show, memory vanishes at late times in a sea of echoes. 
}

\vspace{3.0cm}
\noindent{Essay written for the Gravity Research Foundation 2025 Awards for Essays on Gravitation.}

\noindent{Corresponding email address: Robert.R.Caldwell@Dartmouth.edu}
\vfill\eject

\noindent{\bf Background.}
The concept of gravitational wave ``memory" has long been in use to describe a permanent relative displacement of test masses originating from a change in the source quadrupole moment, within linearized theory \cite{Zeldovich:1974gvh,Braginsky:1985vlg,Braginsky:1987kwo}. Due to the distance of sources and the weakness of gravity, it was generally assumed that higher-order contributions to the memory would be negligible. So it came as some surprise when Christodoulou showed that the nonlinearity of gravitation leads to a non-negligible memory, arising from the stress-energy of the gravitational waves themselves \cite{Christodoulou:1991cr}. Shortly thereafter, Thorne \cite{Thorne:1992sdb} and Wiseman \& Will \cite{Wiseman:1991ss} provided a more physical description and estimates of the memory bursts from binary mergers. Similar results were obtained by Blanchet \& Damour \cite{Blanchet:1992br}. Subsequently, there have been improvements in the theoretical calculations of memory \cite{Favata:2008yd,Talbot:2018sgr}, simulations of waveforms \cite{Favata:2009ii,Mitman:2020bjf,Mitman:2024uss}, detection strategies \cite{McNeill:2017uvq,Xu:2024ybt}, tests of gravity \cite{Jokela:2022rhk,Heisenberg:2023prj},  and forecasts for detection by terrestrial \cite{Lasky:2016knh,Boersma:2020gxx,Hubner:2021amk} and space-based interferometers \cite{Islo:2019qht,Inchauspe:2024ibs,Hou:2024rgo}, as well as pulsar timing \cite{NANOGrav:2023vfo}. New forms of gravitational memory have been discovered \cite{Pasterski:2015tva,Nichols:2017rqr,Flanagan:2018yzh}. And memory has been shown to have deep connections with both the quantum field theory of gravitons and the symmetries of spacetime \cite{Strominger:2014pwa,Kehagias:2016zry,DeLuca:2024bpt}. Today, memory is an active subject and target for discovery \cite{Grant:2022bla}.\\


\noindent{\bf Memory.}
Let us begin by recalling the linearized description of gravitational waves. Consider a perturbation of the spacetime metric $g_{\mu\nu} = \eta_{\mu\nu} + h_{\mu\nu}$ in a Minkowski background. We adopt conventions of a mostly positive spacetime metric and set $c=1$. The Einstein Equations yield the wave equation
\begin{equation}
\Box \bar h_{\mu\nu} = -16 \pi G T_{\mu\nu},
\label{eqn:EE}
\end{equation}
which has general solution
\begin{equation}
    h^{TT}_{ij}(t,\,\vec x) = 4 G \int d^4x' \, \frac{\delta(t-t' - |\vec x - \vec x'|)}{|\vec x - \vec x'|}  T_{ij}(t', \vec x') |^{TT}. \label{eqn:grn}
\end{equation}
Here, $T_{ij}$ is the stress-energy tensor of mechanical sources, such as a binary Keplerian system, and $G$ is Newton's constant. We are justified to use a Minkowski background, provided we restrict attention to time- and length-scales that are small relative to the cosmic expansion scale. We also ignore the Green's function support inside the light cone \cite{Caldwell:1993xw}, which gives rise to gravitational wave tails. To isolate the radiative components, we project out the transverse, traceless portion. This simple equation is sufficient to describe the basic, leading behavior of gravitational waves for a variety of scenarios.

\begin{figure}[t]
    \centering
    \includegraphics[width=1.0\linewidth]{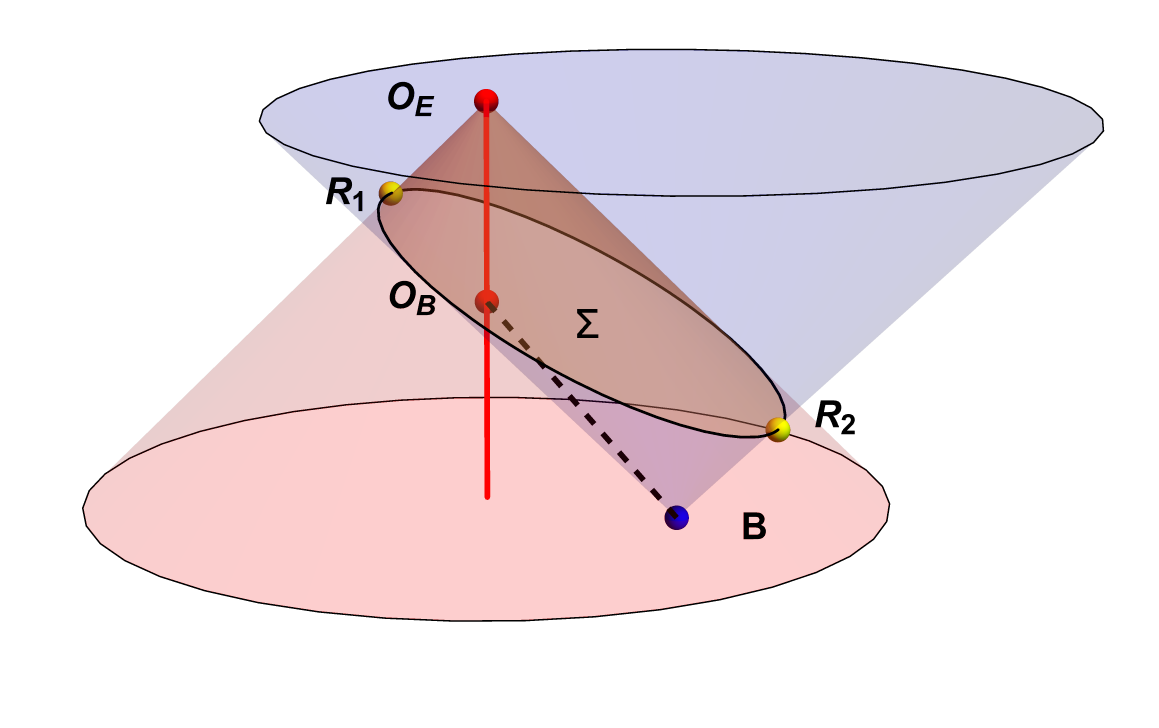}
        \vspace{-2.0cm}\caption{Spacetime diagram illustration of gravitational waves, memory, and subsequent echoes. The observer (thick red line) detects the primary burst of a binary merger (event $B$) at event $O_B$. Intersection of the past light cone of a future observer $O_E$ and the future light cone of the burst is the ellipse, $\Sigma$. Memory of the burst is observed at $O_B$. However, the observer at $O_E$ detects echoes originating on $\Sigma$. Primary waves propagating along $B-R_1$ and $B-R_2$ give rise to secondary waves $R_1-O_E$ and $R_2-O_E$, which are observed as echoes of memory at $O_E$. The null ray along $B-R_1-O_E$ traverses a greater distance than along $B-O_B$, hence the memory at $O_E$ is weaker than at $O_B$.}
    \label{fig:TR}
\end{figure}

The nonlinear Einstein Equations tell us that the energy conveyed by gravitational waves is itself the source of further gravitational waves \cite{Christodoulou:1991cr}. That is, the stress-energy tensor of the primary gravitational waves also contributes to the right hand side of Eq.~(\ref{eqn:EE}). Treated iteratively, $h \to h + \delta h$,
\begin{equation}
    \delta h^{TT}_{ij}(t,\,\vec x) = 4 G \int d^4x' \, \frac{\delta(t-t' - |\vec x - \vec x'|)}{|\vec x - \vec x'|} T^{GW}_{ij}(t',\,\vec x')\, |^{TT} ,\label{eqn:TT}
\end{equation}
where $n$ are outward directed unit vectors pointing along the direction of radiation, and
\begin{equation}
T^{GW}_{ij} = \frac{1}{r^2}\frac{dL}{d\hat n} n_i n_j \qquad 
   \frac{1}{r^2} \frac{dL}{d\hat n} = \frac{1}{32 \pi G} \partial_t h^{TT}_{ij} \partial_t h^{ij}_{TT}
\end{equation}
where $L$ is the luminosity of the primary gravitational waves. This insightful model was first presented in Refs.~\cite{Wiseman:1991ss,Thorne:1992sdb}. A diagram illustrating the origin and propagation of the primary and secondary gravitational waves is shown in Fig.~\ref{fig:TR}.

The expression for the secondary gravitational wave may be simplified by introducing the delayed time $u \equiv t-r$, whereby
\begin{equation}
    \delta h^{TT}_{ij}  = 4 G \int du' \, d\hat n' \, \frac{dL}{d\hat n'} \left[ \frac{n'_i n'_j}{t-u' - \vec x \cdot \hat n'} \right]|^{TT} .\label{eqn:general}
\end{equation}
Restoring $c$ for a moment, the right hand side is proportional to $c^{-5}$, which we naively interpret to mean $\delta h^{TT}$ is of order 2.5 in the post-Newtonian perturbative expansion. Yet the secondary source enters the waveform at the same order as the primary; one wave sourced by mechanical stress, the other by energy.

The above expression will be of use for describing the gravitational waves along the world line of an observer, $O$, at a time $t$ and a distance $r$ from the source. We can infer from the above expression that $\delta h$ accumulates monotonically as long as the original mechanical system continues to radiate -- through inspiral, merger, and ringdown for a binary system. After ringdown, $\delta h$ ceases to grow, and the observer records a finite, residual displacement of test masses. This is the origin of the nonlinear gravitational wave memory. But we further note that at late times, $\delta h$ vanishes as $t \to \infty$. Memory is not forever.\\

\noindent{\bf Short-term memory.}
Now consider the secondary gravitational wave, $\delta h^{TT}_{ij}$, resulting from a burst of primary gravitational waves arising from a source located at a distance $r$ from the observer. For a time interval $\Delta t$ much shorter than the light travel time of the burst from source to observer, then $t-u' \simeq r$, in which case
\begin{equation}
    \delta h^{TT}_{ij}  = \frac{4 G}{r} \int du' \, d\hat n' \, \frac{dL}{d\hat n'} \left[ \frac{n'_i n'_j}{1 - \hat n \cdot \hat n'} \right]|^{TT} .\label{eqn:static}
\end{equation}
This is the familiar, general expression for nonlinear gravitational wave memory. It is static, but only valid for short time intervals, centered on event $O_B$ in Fig.~\ref{fig:TR}. We will refer to $\delta h$ from Eq.~(\ref{eqn:static}) as the short-term memory.

Consider the ideal case of radiation from binary inspiral. We will make the simplifying assumption that the primary waveform is dominated by the quadrupolar amplitude, 
\begin{equation}
    h \equiv h_+ - i h_\times = \sum_{\ell=2}^\infty \sum_{m=-\ell}^\ell\, h_{\ell m}(t,r)\, {}_{-2}Y_{\ell,\, m}(\theta,\,\phi) \simeq \sum_{m=\pm 2}h_{2m}(t,r)\, {}_{-2}Y_{2,\, m}(\theta,\,\phi),
\end{equation} 
where $h_{22}$ is complex and ${}_{-2}Y_{\ell m}$ are spin-2 spherical harmonics. In this case, Eq.~(\ref{eqn:static}) may be written
\begin{equation}
    r \delta h^{TT} = \frac{1}{8 \pi} \int du' \, r^2 |\dot h^{TT}_{22}|^2 \int d\hat n' \, \frac{Q(\hat n,\, \hat n')}{1 - \hat n \cdot \hat n'}, 
\end{equation}
where an oscillatory cross term is assumed to vanish. Here, $Q=\left(|{}_{-2} Y_{2,\,2}(\hat n')|^2 + |{}_{-2} Y_{2,\,-2}(\hat n')|^2\right)$ $\times$  ${n_a' n_b'}\Pi_{abij}(\hat n)(e_+^{ij}(\hat n) - i e_\times^{ij}(\hat n))$, $\Pi_{abij} = P_{ai} P_{bj} - \frac{1}{2} P_{ab} P_{ij}$, $P_{ab} = \delta_{ab} - n_a n_b$ is the transverse projection tensor, and $e^{ij}_{+,\times}$ are the polarization tensors, transverse to $n$. The angular integration is straight forward, and we arrive at the result
\begin{equation}
    r \delta h^{TT} = -\frac{1}{192 \pi} (17 + \cos^2\theta)\sin^2\theta \int du' \, r^2 |\dot h^{TT}_{22}|^2  \label{eqn:memgen}
\end{equation}
for the memory due to any quadrupole-dominated primary waveform.

\begin{figure}[b]
    \centering
    \includegraphics[width=0.75\linewidth]{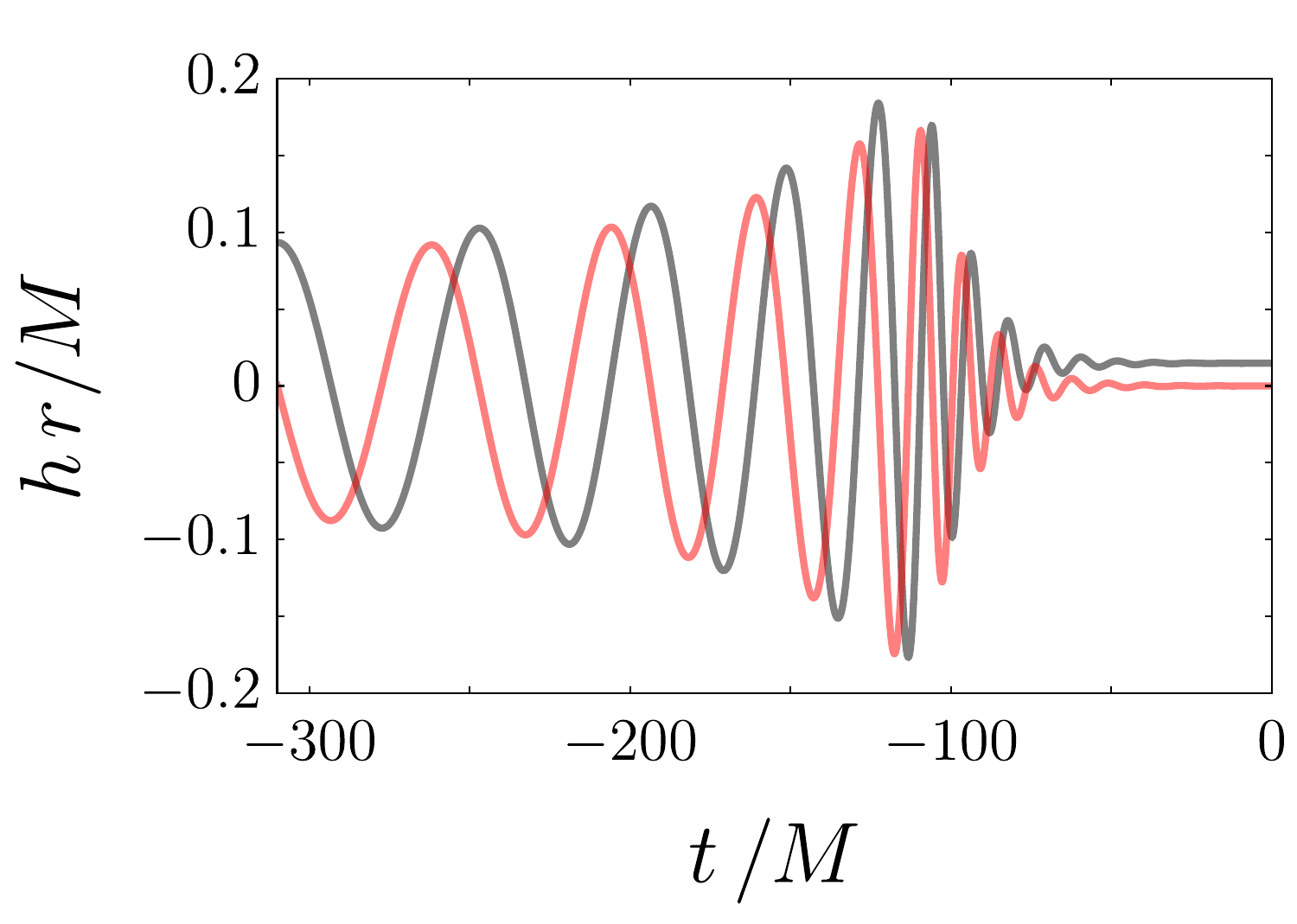}
        \caption{The $+$ (black) and $\times$ (red) polarization waveforms including memory are illustrated for an equal mass, non-spinning, black hole binary at inclination $\pi/4$. The offset of the $+$ waveform as $t\to 0$ is due to the memory.}
    \label{fig:waveforms}
\end{figure}

In the case of binary inspiral of a circularized, equal mass Keplerian system, $|\dot h^{TT}_{22}|^2 = {256 \pi}(\pi f G {\cal M})^{10/3}/{5 r^2}$, where ${\cal M}$ is the chirp mass and $f$ is the wave frequency. To evaluate Eq.~(\ref{eqn:memgen}), the integral over delayed time may be converted into an integral over frequency. The cumulative memory amplitude is
\begin{equation}
    r \delta h^{TT} = -\frac{1}{48} (\pi f_I)^{2/3} (G {\cal M})^{5/3} (17 + \cos^2\theta) \sin^2\theta, \label{eqn:memory}
\end{equation}
where $f_I$ is the radiation frequency at the upper bound of the delayed-time integration, which we take to be the frequency at the innermost stable circular orbit. The above expression agrees with the standard result \cite{Favata:2008yd,Favata:2009ii}. We note that the sign of the displacement is negative, a consequence of how we define our polarization basis; by rotating the basis vectors $\pi/2$ we may change the sign of Eq.~(\ref{eqn:memgen}), which we do. Higher moments beyond the quadrupole approximation are required to remove the sign ambiguity relative to the primary \cite{Lasky:2016knh}.

For comparison, the rms amplitude of the primary gravitational wave is
\begin{equation}
    r h^{TT}_{rms} =  \sqrt{2} (\pi f)^{2/3}(G {\cal M})^{5/3} \sqrt{1 + 6 \cos^2\theta + \cos^4\theta}.
\end{equation}
For the purposes of comparison, we may evaluate the primary waveform amplitude at $f_I$. In this case, for inclination $\pi/4$, $|\delta h^{TT}|/{h^{TT}_{rms}} \simeq 0.06$, which is roughly in accord with numerical results \cite{Favata:2009ii,Lasky:2016knh}. 

To obtain a more realistic estimate of the nonlinear GW memory, we can evaluate Eq.~(\ref{eqn:memgen}) using a numerical, primary waveform \cite{Varma:2018mmi}, as illustrated in Fig.~{\ref{fig:waveforms}. The nonlinear memory displacement is easy to identify.

\begin{figure}[b]
    \centering
    \includegraphics[width=0.75\linewidth]{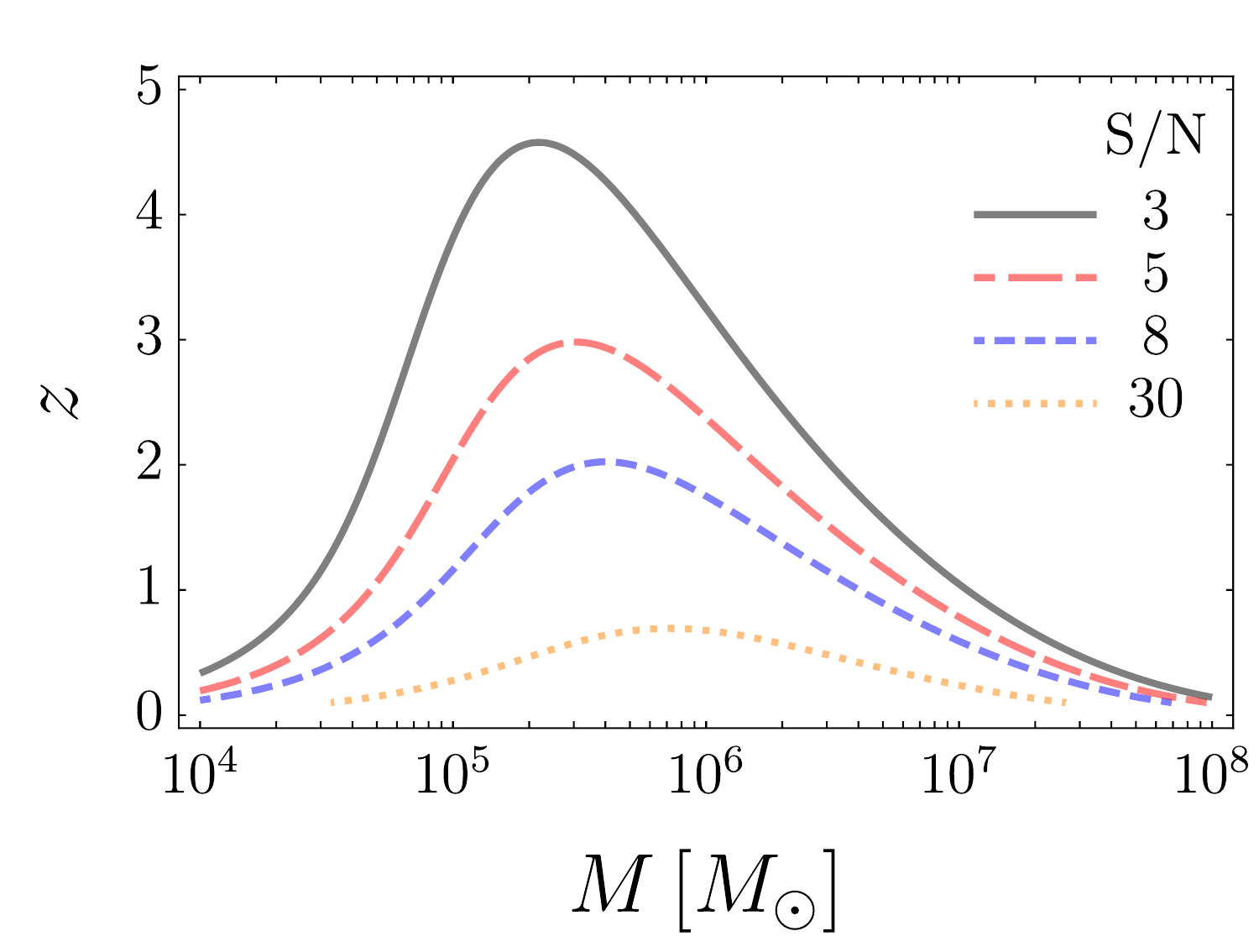}
        \caption{Contours of constant signal-to-noise ratio for LISA to detect the nonlinear GW memory from the merger of equal mass, non-spinning black holes of mass $M$ at redshift $z$.}
    \label{fig:sensitivity}
\end{figure}

Thus encouraged, we may ask if the memory is detectable. To be specific, we consider the sensitivity of the Laser Interferometric Space antenna (LISA) to the memory in the simple case of non-spinning, equal mass black hole binaries. We use Eqs.~(86-87) of Ref.~\cite{Smith:2019wny} to evaluate the signal-to-noise ratio, for different masses and redshifts, averaging over inclination angles. The results  illustrated in Fig.~\ref{fig:sensitivity} suggest that, given the existence of sources within cosmological distances, the detection of nonlinear memory should be feasible.\\


\noindent{\bf Long-term memory.}
Now let us examine the long-term memory. If we drop the assumption of a short time interval after arrival of the main burst, then the memory for a quadrupole-dominated source is given by
\begin{equation}
    r \delta h^{TT} = \frac{1}{8 \pi} \int du' \, r^2 |\dot h^{TT}_{22}|^2 \int d\hat n' \frac{Q(\hat n,\, \hat n')}{\frac{t-u'}{r} - \hat n \cdot \hat n'}.
\end{equation}
Since the time-scale for binary merger is much smaller than the light travel time to the source, we are justified to describe the primary waveform as a burst that is concentrated at $u'=0$, e.g.
\begin{equation}
    | \dot h_{22}^{TT}|^2 = \frac{B_{22}}{r^2} \delta(u')
\end{equation}
(where $B$ is for ``burst"). We infer that $B_{22} = \int du' \, r^2 |\dot h_{22}^{TT}|^2$ in general, whereas for the particular case of the Keplerian binary, $B_{22} = 4 \pi (\pi f)^{2/3}(G {\cal M})^{5/3}$. In the general case, the memory reduces to  
\begin{equation}
    r \delta h^{TT} = \frac{B_{22}}{8 \pi} \int d\hat n' \, \frac{Q(\hat n,\, \hat n')}{s - \hat n \cdot \hat n'}
\end{equation}
where $s \equiv t/r \ge 1$. The angular integrals may be evaluated after some work. The  final expression is 
\begin{equation}
    r \delta h^{TT} = \frac{B_{22}}{8 \pi} \times \frac{1}{192} \left[ \left(C_1 + C_2 \ln\frac{s-1}{s+1}\right) + \cos^2\theta \left( C_3 + C_4 \ln\frac{s-1}{s+1}\right) \right] \sin^2\theta  \label{eqn:persist}
\end{equation}
where $C_1 =  -2 s(183 - 130 s^2 + 15 s^4)$, $C_2 = -15(s^2-7)(s^2-1)^2$, $C_3 = 2 s(81 - 190 s^2 + 105 s^4)$, and $C_4 = 15 (7 s^2-1)(s^2-1)^2$ for $s = t/r \ge 1$. Eq.~(\ref{eqn:persist}) is our main result. 

Eq.~(\ref{eqn:persist}) has the correct, limiting behavior. First, in the limit $s \to 1$, it simplifies to the standard result for $\Delta t \ll r$,
\begin{equation}
    r \delta h^{TT}|_{s=1} = -\frac{B_{22}}{192\pi}(17 + \cos^2\theta) \sin^2\theta.
\end{equation}
Second, in the limit $s \to \infty$, corresponding to $\Delta t \gg r$, 
\begin{equation}
    r \delta h^{TT} \approx -\frac{B_{22}}{14\pi} s^{-1} \sin^2\theta + {\cal O}(s^{-2}).
\end{equation}
That is, the nonlinear memory observed by $O$ ultimately vanishes.

We note that the expression for the global memory evaluated at future null infinity takes the same form as the short-term memory, since $s\to 1$. There is no contradiction, however: while the memory at any particular location vanishes as $t,\, r \to \infty$, the net power is finite.\\


\noindent{\bf Persistence.}
To examine the decline of memory, we define
\begin{equation}
    r \delta h^{TT} \equiv \left( r \delta h^{TT} |_{s=1} \right) \times P(s,\,\theta)
\end{equation}
where $P$ is the {\it persistence} function. As illustrated in Fig.~\ref{fig:persistence}, $P(s,\, \theta)$ describes the long-time behavior of the nonlinear gravitational wave memory. Using the leading term for $P$ for $s \to \infty$, we can make the following approximation:
\begin{equation}
    P \approx \frac{96}{7 (17 + \cos^2\theta) s }.
\end{equation}
As illustrated in Fig.~\ref{fig:persistence}, $P \approx 0.8/s$ in the case $\theta = \pi/2$.

\begin{figure}[b]
    \centering
    \includegraphics[width=0.75\linewidth]{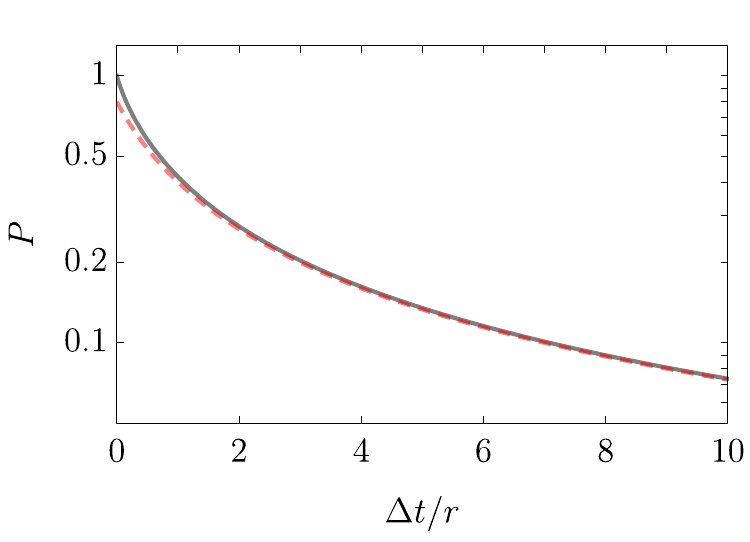}
        \caption{Nonlinear gravitational wave memory persistence $P$ (solid, black) versus time since burst, $\Delta t$, in units of distance to source, $r$. An edge-on alignment, $\theta = \pi/2$, is assumed. A late-time approximation $P \propto (1 + \Delta t/r)^{-1}$ (red, dashed) provides a good description of the decline of memory.}
    \label{fig:persistence}
\end{figure}

To an observer along $O$ in Fig.~\ref{fig:TR}, the arrival of the primary burst is coincident with the accumulation of the nonlinear memory, a residual offset of the detector test masses. After passage of time comparable to the light-travel time to the source, the observer will find that the memory displacement has decayed. To put in some numbers, LIGO now regularly detects binary BH mergers at luminosity distances $\sim 1$~Gpc; in the standard cosmology, these sources are at redshift $z\sim 0.2$, corresponding to a light-travel time of $\sim 2.5$~Gyr.

If a late observer misses the burst waveform, we can imagine the albeit impractical situation whereby they may still turn on their detector at $s_{on}>1$. They will receive the decaying signal, which will appear as a slowly-accumulating waveform
\begin{equation}
    \Delta h^{TT}(s) =\delta h^{TT}(s) - \delta h^{TT}(s_{on}),
\end{equation}
which asymptotes to $\Delta h^{TT} \to - \delta h^{TT}(s_{on})$. Hence, the late-arriving gravitational waves may be thought of as memory echoes. For the early observer, these echoes erode the original nonlinear memory displacement; for the late observer, missing the burst, the subsequent echoes accumulate into a new memory displacement of equal magnitude to that deposited by the burst.

The memory echo waveform decays with time as $s^{-1}$, so its Fourier transform breaks at $f r \sim 1$. Hence, the decay waveform is quite the opposite of a chirp or burst. Rather, the waveform is a dim, low-frequency chorus at $f \lesssim r^{-1}$.

The collective memory echoes of astrophysical sources comprise a low-level, low frequency gravitational wave background. It is straight-forward to compute the spectral density
\begin{equation}
    \Omega_{GW}(f) = \frac{1}{\rho_c}\int dz \, \frac{n(z)}{1+z} \frac{dE_{GW}}{d\ln f},
\end{equation}
where $\rho_c$ is the critical density, $n$ is the number density of gravitational wave sources, and $E_{GW}$ is the energy released by such a source \cite{Phinney:2001di}. These sources are bursts that lie within the observer's past light cone, creating echoes on the past light cone. Owing to the very low frequencies of the echoes, the energy density is very weak, like an imperceptible fog. The Universe is awash in a variety of sources of such long-wavelength gravitational waves.


In short, we have shown that memory does not last; the memory of a burst declines in a sea of echoes; patient observers who only receive the echoes can reconstruct the missing, original memory.

\vfill\eject

\bibliography{main}


\vfill\eject

\end{document}